\documentclass[12pt]{iopart}
\usepackage{bm,amsfonts,amssymb, graphicx}
\usepackage{cite}
\usepackage{epstopdf}

\begin{document}

\title[Harmonic generation in topological insulator]{Multiphoton excitation
and high-harmonics generation in topological insulator}
\author{H K Avetissian, A K Avetissian, B R Avchyan and G F Mkrtchian}

\address{Centre of Strong Fields Physics, Yerevan State University, 1 A. Manukian,
Yerevan 0025, Armenia}

\begin{abstract}
Multiphoton interaction of coherent electromagnetic radiation with 2D
metallic carriers confined on the surface of the 3D topological insulator is
considered. A microscopic theory describing the nonlinear interaction of a
strong wave and metallic carriers with many-body Coulomb interaction is
developed. The set of integrodifferential equations for the interband
polarization and carrier occupation distribution is solved numerically.
Multiphoton excitation of Fermi-Dirac sea of 2D massless carriers is
considered for a THz pump wave. It is shown that in the moderately strong
pump wave field along with multiphoton interband/intraband transitions the
intense radiation of high harmonics takes place.
\end{abstract}

\maketitle



\section{ Introduction}

Along with graphene topological insulators (TI) recently emerged as a
central theme in condensed matter physics \cite{TIR1,TIR2}.
Three-dimensional TIs are bulk insulators endowed with a topological
invariant that manifests itself through robust 2D metallic surface states.
These states are helical with massless linear Dirac like energy dispersion,
that is, each surface-momentum state possesses a unique spin direction and
are protected against backscattering by time-reversal symmetry \cite%
{exp1,exp2}. The unique properties of the surface states are responsible for
their exotic electromagnetic properties. Several theoretical works on
light-TI interaction have illustrated interesting effects \cite%
{TIL2,TIL3,TIL4,PhotD,Magnetao}. Experimentally, Kerr \cite{Kerr} and
Faraday \cite{Farad} effects, and second-harmonic generation \cite{2nd}
process in TI have been studied. Metallic surface states being Dirac like
are responsible for strong nonlinear terahertz response of TI \ \cite%
{DiracM,THzpl,THzStrong}, and like to graphene TIs have great potential as
an effective nonlinear optical material \cite{HH2}. In particular, in the
strong pump field limit, one can realize the regime where multiphoton
effects are essential \cite{Mer1,Mer2,Mer3} and high-harmonics are
generated. The experiment \cite{HH3-exp} with the generation of ninth
harmonic in graphene opens new avenue towards high-harmonic generation in 2D
nanostructures. Hence it is of interest to investigate multiphoton
excitation and subsequent high harmonic generation process in TIs.

In the present work, we develop a nonlinear microscopic quantum theory of
interaction of 2D metallic carriers confined on the surface of the 3D TI
(e.g. Bi$_{2}$Se$_{3}$) with coherent electromagnetic radiation. We also
take into account electron-electron Coulomb interaction with induced
many-body effects. We consider nonlinear coherent interaction in the
ultrafast excitation regime when relaxation processes due to electron-phonon
coupling via longitudinal surface phonons are not relevant. We use the
self-consistent Hartree-Fock approximation that leads to a closed set of
integrodifferential equations for the interband polarization and carrier
occupation distribution. The latter is solved numerically. Then we consider
high harmonic generation process for strong pump waves and show that one can
achieve efficient generation of high harmonics in TIs.

The paper is organized as follows. In Sec. II the set of equations for the
interband polarization and carrier occupation distribution is formulated. In
Sec. III, we consider multiphoton excitation of Fermi-Dirac sea and
generation of harmonics in TI. Finally, conclusions are given in Sec. IV.

\section{Evolutionary equation for single-particle density matrix}

Low-energy excitations of 2D metallic surface states of TI which are much
smaller than the bulk gap energy ($0.3\ \mathrm{eV}$ for $\mathrm{Bi}_{2}%
\mathrm{Se}_{3}$) can be described by the effective Hamiltonian

\begin{equation}
H_{0}=\hbar \mathrm{v}_{F}\left( k_{x}\sigma _{y}-k_{y}\sigma _{x}\right)
=i\hbar \mathrm{v}_{F}\left( 
\begin{array}{cc}
0 & -k_{x}+ik_{y} \\ 
k_{x}+ik_{y} & 0%
\end{array}%
\right) ,  \label{DH}
\end{equation}%
where $\mathrm{v}_{F}\approx c/450$ is the Fermi velocity for the
topological insulator ($c$ is the light speed in vacuum),\textbf{\ }$\hbar 
\mathbf{k}$\textbf{\ }is the 2D electron momentum operator. The Pauli
matrices $\sigma _{x}$ and $\sigma _{y}$ act in the real electron spin
space. The eigenstates of the effective Hamiltonian (\ref{DH}) are

\begin{equation}
|\psi _{\eta ,\mathbf{k}}(\mathbf{r})\rangle =\frac{1}{\sqrt{\mathcal{A}}}%
|\varphi _{\eta ,\mathbf{k}}\rangle e^{i\mathbf{kr}},  \label{free}
\end{equation}%
where the spinors%
\[
|\varphi _{\eta ,\mathbf{k}}\rangle =\frac{1}{\sqrt{2}}\left( 
\begin{array}{c}
e^{-i\theta \left( \mathbf{k}\right) } \\ 
i\eta%
\end{array}%
\right) 
\]%
correspond to energies%
\[
\mathcal{E}_{\eta }\left( \mathbf{k}\right) =\eta \hbar \mathrm{v}_{F}k. 
\]%
Here for conduction band $\eta =1$ and for valence band $\eta =-1$, $%
\mathcal{A}$ is the surface area, and 
\begin{equation}
\theta \left( \mathbf{k}\right) =\arctan \left( \frac{k_{y}}{k_{x}}\right)
\label{angle}
\end{equation}%
is the polar angle in the momentum space. The mean value of an electron spin
in the 2D surface states of TI is

\begin{equation}
\frac{\hbar }{2}\langle \psi _{\eta ,\mathbf{k}}|\mathbf{\sigma }|\psi
_{\eta ,\mathbf{k}}\rangle =\frac{\eta \hbar }{2}\left[ \widehat{\mathbf{z}}%
\times \widehat{\mathbf{k}}\right] ,  \label{spin}
\end{equation}%
where $\widehat{\mathbf{k}}$ and $\widehat{\mathbf{z}}$ are unit vectors
directed along the $\mathbf{k}$ and the $z$-axis (normal to the surface),
respectively. As is seen from Eq. (\ref{spin}), in TI the spin of electron
lies in the surface plane and is perpendicular to its momentum. At that, for
conduction band it is directed in the counterclockwise direction and
inversely for the valence band.

Let a plane linearly polarized (along the $x$-axis) quasimonochromatic
electromagnetic radiation of carrier frequency $\omega _{0}$ and slowly
varying envelope $E_{0}(t)$ interacts with the 3D TI. We assume
perpendicular to the metallic surface incidence and $\hbar \omega _{0}<<%
\mathcal{E}_{g}$ ($\mathcal{E}_{g}$ is the TI's bulk gap). Besides, we will
restrict wave intensity to forbidden transition within the bulk bands. Under
these circumstances, one can neglect bulk excitations and the nonlinear
electromagnetic response of TI will be conditioned by the 2D surface states.
Thus, the light--TI interaction Hamiltonian in the length gauge will be: 
\begin{equation}
\widehat{H}_{\mathrm{int}}=e\int d\mathbf{r}\ \widehat{\Psi }^{\dagger }(%
\mathbf{r})\mathbf{rE}\left( t\right) \widehat{\Psi }(\mathbf{r}),
\label{Eint}
\end{equation}%
where $\widehat{\Psi }(\mathbf{r})$ is the fermionic field operator and 
\begin{equation}
\mathbf{E}\left( t\right) =\widehat{\mathbf{x}}E_{0}\left( t\right) \cos
\omega _{0}t.  \label{E_field}
\end{equation}%
We will work in the second quantization formalism using the Fermi-Dirac
field operator%
\begin{equation}
\widehat{\Psi }(\mathbf{r})=\sum\limits_{\mathbf{k,}\eta }\widehat{e}_{\eta ,%
\mathbf{k}}|\psi _{\eta ,\mathbf{k}}(\mathbf{r})\rangle ,  \label{exp}
\end{equation}%
where $\widehat{e}_{\eta \lambda ,\mathbf{k}}$ ($\widehat{e}_{\eta ,\mathbf{k%
}}^{\dagger }$) is the annihilation (creation) operator for an electron with
momentum $\mathbf{k}$ and band $\eta =\pm 1$.

The electrons interact through the long-range Coulomb forces and the
Hamiltonian for electron-electron interactions can be written in terms of
the field operators $\widehat{\Psi }(\mathbf{r})$, as:%
\[
\widehat{H}_{\mathrm{c}}=\frac{1}{2}\int d\mathbf{r}\int d\mathbf{r}^{\prime
}\ \widehat{\Psi }^{\dagger }(\mathbf{r})\widehat{\Psi }^{\dagger }(\mathbf{r%
}^{\prime })V_{c}(\mathbf{r}-\mathbf{r}^{\prime })\widehat{\Psi }(\mathbf{r}%
^{\prime })\widehat{\Psi }(\mathbf{r}), 
\]%
where $V_{c}(\mathbf{r})=e^{2}/\left( \varepsilon \left\vert \mathbf{r}%
\right\vert \right) $ is the bare Coulomb potential, $\varepsilon $ is the
effective dielectric constant of the TI.

Taking into account expansion (\ref{exp}), the total Hamiltonian can be
represented as follow: 
\[
\widehat{H}=\sum_{\eta ,\mathbf{k}}\mathcal{E}_{\eta }\left( \mathbf{k}%
\right) \widehat{e}_{\eta ,\mathbf{k}}^{\dag }\widehat{e}_{\eta ,\mathbf{k}%
}+H_{\mathrm{Coul}}+\frac{e\mathbf{E}\left( t\right) }{\mathcal{A}} 
\]%
\begin{equation}
\times \sum\limits_{\eta ,\eta ^{\prime }}\sum\limits_{\mathbf{k,k}^{\prime
}}\int \mathbf{r}d\mathbf{r}e^{i\left( \mathbf{k-k}^{\prime }\right) \mathbf{%
r}}\ \langle \varphi _{\eta ^{\prime },\mathbf{k}^{\prime }}||\varphi _{\eta
,\mathbf{k}}\rangle \widehat{e}_{\eta ^{\prime },\mathbf{k}^{\prime
}}^{\dagger }\widehat{e}_{\eta ,\mathbf{k}}.  \label{Ham12}
\end{equation}%
The Coulomb interaction reads: 
\[
H_{\mathrm{Coul}}=\frac{1}{2\mathcal{A}}\sum_{\eta _{1}\eta _{2}}\sum_{\eta
_{3}\eta _{4}}\sum_{\mathbf{q,k,k}^{\prime }}V_{2D}\left( \mathbf{q}\right)
\digamma _{\eta _{1}\eta _{2}\eta _{3}\eta _{4}}\left( \mathbf{q,k},\mathbf{k%
}^{\prime }\right) 
\]%
\begin{equation}
\times \widehat{e}_{\eta _{1},\mathbf{k+q}}^{\dag }\widehat{e}_{\eta _{2},%
\mathbf{k}^{\prime }-\mathbf{q}}^{\dag }\widehat{e}_{\eta _{3},\mathbf{k}%
^{\prime }}\widehat{e}_{\eta _{4},\mathbf{k}},  \label{CB}
\end{equation}%
where%
\begin{equation}
V_{2D}\left( \mathbf{q}\right) =\frac{2\pi e^{2}}{\varepsilon \left\vert 
\mathbf{q}\right\vert }  \label{2DC}
\end{equation}%
is the 2D Coulomb potential in the momentum space and 
\begin{equation}
\digamma _{\eta _{1}\eta _{2}\eta _{3}\eta _{4}}\left( \mathbf{q,k},\mathbf{k%
}^{\prime }\right) =\langle \varphi _{\eta _{1},\mathbf{k+q}}||\varphi
_{\eta _{4},\mathbf{k}}\rangle \langle \varphi _{\eta _{2},\mathbf{k}%
^{\prime }-\mathbf{q}}||\varphi _{\eta _{3},\mathbf{k}^{\prime }}\rangle .
\label{form}
\end{equation}%
In the light--TI interaction part of the Hamiltonian (\ref{Ham12}) there are
terms responsible for intraband transitions ($\eta =\eta ^{\prime }$), as
well as terms that describe interband transitions ($\eta =-\eta ^{\prime }$).

In order to develop a microscopic theory of the multiphoton interaction of
TI with a strong radiation field, we need to solve the Liouville--von
Neumann evolution equation for a single-particle density matrix,%
\begin{equation}
\rho _{\eta _{1},\eta _{2}}(\mathbf{k}_{1},\mathbf{k}_{2},t)=\langle 
\widehat{e}_{\eta _{2},\mathbf{k}_{2}}^{\dag }\left( t\right) \widehat{e}%
_{\eta _{1},\mathbf{k}_{1}}\left( t\right) \rangle ,  \label{def}
\end{equation}%
\qquad\ where $\widehat{e}_{\eta ,\mathbf{k}}\left( t\right) $ obeys the
Heisenberg equation%
\begin{equation}
i\hbar \frac{\partial \widehat{e}_{\eta ,\mathbf{k}}\left( t\right) }{%
\partial t}=\left[ \widehat{e}_{\eta ,\mathbf{k}}\left( t\right) ,\widehat{H}%
\right] ,  \label{Heiz}
\end{equation}%
and expectation values are determined by the initial density matrix. Due to
the homogeneity of the problem, we only need the $\mathbf{k}$-diagonal
elements of the density matrix. The $\mathbf{k}$-diagonal elements represent
particle distribution functions for conduction $N_{c}\left( \mathbf{k}%
,t\right) =\rho _{1,1}(\mathbf{k},\mathbf{k},t)$ and for valence $%
N_{v}\left( \mathbf{k},t\right) =\rho _{-1,-1}(\mathbf{k},\mathbf{k},t)$
bands, and interband polarization $P\left( \mathbf{k},t\right) =\rho _{-1,1}(%
\mathbf{k},\mathbf{k},t)=\rho _{1,-1}^{\ast }(\mathbf{k},\mathbf{k},t)$. We
just need equations for $N_{c}\left( \mathbf{k},t\right) $, $N_{v}\left( 
\mathbf{k},t\right) $ and $P\left( \mathbf{k},t\right) $. The Coulomb
interaction part (\ref{CB}) contains products of four fermionic operators.
For the closed set of equations, we need to reduce it into products of two
fermionic operators. Thus, Coulomb interaction we will treat under
Hartree-Fock approximation, which is valid for short time scales. The
Hartree contribution $\sim V_{2D}\left( \mathbf{q=0}\right) $ is zero, which
is physically related to the neutrality of charge of the total system. For
the Fock part we will use decomposition:%
\[
\widehat{e}_{\eta _{1},\mathbf{k+q}}^{\dag }\widehat{e}_{\eta _{2},\mathbf{k}%
^{\prime }-\mathbf{q}}^{\dag }\widehat{e}_{\eta _{3},\mathbf{k}^{\prime }}%
\widehat{e}_{\eta _{4},\mathbf{k}}=-\left\{ \widehat{e}_{\eta _{1},\mathbf{k}%
^{\prime }}^{\dag }\widehat{e}_{\eta _{3},\mathbf{k}^{\prime }}\langle 
\widehat{e}_{\eta _{2},\mathbf{k}}^{\dag }\widehat{e}_{\eta _{4},\mathbf{k}%
}\rangle \right. 
\]%
\begin{equation}
\left. +\widehat{e}_{\eta _{2},\mathbf{k}}^{\dag }\widehat{e}_{\eta _{4},%
\mathbf{k}}\,\langle \widehat{e}_{\eta _{1},\mathbf{k}^{\prime }}^{\dag }%
\widehat{e}_{\eta _{3},\mathbf{k}^{\prime }}\rangle -\langle \,\widehat{e}%
_{\eta _{1},\mathbf{k}^{\prime }}^{\dag }\widehat{e}_{\eta _{3},\mathbf{k}%
^{\prime }}\rangle \langle \widehat{e}_{\eta _{2},\mathbf{k}}^{\dag }%
\widehat{e}_{\eta _{4},\mathbf{k}}\rangle \right\} \delta _{\mathbf{q,k}%
^{\prime }-\mathbf{k}}.  \label{Fock}
\end{equation}%
Taking into account definition (\ref{def}), the second quantized Hamiltonian
(\ref{Ham12}), and Eqs. (\ref{CB}, \ref{Fock}), from Eq. (\ref{Heiz}) one
can obtain the following equations for $N_{c}\left( \mathbf{k},t\right) $, $%
N_{v}\left( \mathbf{k},t\right) $ and $P\left( \mathbf{k},t\right) $:%
\[
\frac{\partial N_{c}\left( \mathbf{k},t\right) }{\partial t}-\frac{e\mathbf{E%
}}{\hbar }\frac{\partial N_{c}\left( \mathbf{k},t\right) }{\partial \mathbf{k%
}} 
\]%
\begin{equation}
=i\left( \Omega _{R}\left( \mathbf{k},t\right) +\Omega _{\mathrm{PN}}\left( 
\mathbf{k},t\right) \right) P^{\ast }\left( \mathbf{k},t\right) +\mathrm{c.c.%
},  \label{ev1}
\end{equation}%
\[
\frac{\partial N_{v}\left( \mathbf{k},t\right) }{\partial t}-\frac{e\mathbf{E%
}}{\hbar }\frac{\partial N_{v}\left( \mathbf{k},t\right) }{\partial \mathbf{k%
}} 
\]%
\begin{equation}
=-i\left( \Omega _{R}\left( \mathbf{k},t\right) +\Omega _{\mathrm{PN}}\left( 
\mathbf{k},t\right) \right) P^{\ast }\left( \mathbf{k},t\right) +\mathrm{c.c.%
},  \label{ev2}
\end{equation}%
\[
\frac{\partial P\left( \mathbf{k},t\right) }{\partial t}-\frac{e\mathbf{E}}{%
\hbar }\frac{\partial P\left( \mathbf{k},t\right) }{\partial \mathbf{k}}=i%
\left[ \omega _{0}\left( \mathbf{k}\right) +\omega _{\mathrm{PN}}\left( 
\mathbf{k},t\right) \right] P\left( \mathbf{k},t\right) 
\]%
\begin{equation}
-i\left( \Omega _{R}\left( \mathbf{k},t\right) +\Omega _{\mathrm{PN}}\left( 
\mathbf{k},t\right) \right) \left( N_{c}\left( \mathbf{k},t\right)
-N_{v}\left( \mathbf{k},t\right) \right) ,  \label{ev3}
\end{equation}%
where%
\begin{equation}
\Omega _{R}\left( \mathbf{k},t\right) =\frac{e\mathbf{E}}{2\hbar }\frac{%
\partial \theta (\mathbf{k})}{\partial \mathbf{k}}  \label{Rabi}
\end{equation}%
is the Rabi frequency and 
\[
\Omega _{\mathrm{PN}}\left( \mathbf{k},t\right) =-i\frac{1}{2\hbar \mathcal{A%
}}\sum_{\mathbf{k}^{\prime }\neq \mathbf{k}}V_{2D}\left( \mathbf{k}-\mathbf{k%
}^{\prime }\right) \sin [\theta (\mathbf{k})-\theta (\mathbf{k}^{\prime })] 
\]%
\[
\times \left( N_{c}\left( \mathbf{k}^{\prime },t\right) -N_{v}\left( \mathbf{%
k}^{\prime },t\right) \right) -\frac{1}{\hbar \mathcal{A}}\sum_{\mathbf{k}%
^{\prime }\neq \mathbf{k}}V_{2D}\left( \mathbf{k}-\mathbf{k}^{\prime
}\right) 
\]%
\begin{equation}
\times \left[ P^{\prime }\left( \mathbf{k}^{\prime },t\right) +i\cos [\theta
(\mathbf{k})-\theta (\mathbf{k}^{\prime })]P^{\prime \prime }\left( \mathbf{k%
}^{\prime },t\right) \right] .  \label{NPRabi}
\end{equation}%
The transition frequency is defined by 
\begin{equation}
\omega \left( \mathbf{k}\right) =2\mathrm{v}_{F}k  \label{trF}
\end{equation}%
and%
\[
\omega _{\mathrm{PN}}\left( \mathbf{k},t\right) =\frac{1}{\hbar \mathcal{A}}%
\sum_{\mathbf{k}^{\prime }\neq \mathbf{k}}V_{2D}\left( \mathbf{k}-\mathbf{k}%
^{\prime }\right) 
\]%
\[
\times \cos [\theta (\mathbf{k})-\theta (\mathbf{k}^{\prime })]\left(
N_{v}\left( \mathbf{k}^{\prime },t\right) -N_{c}\left( \mathbf{k}^{\prime
},t\right) \right) 
\]%
\begin{equation}
+\frac{2}{\hbar \mathcal{A}}\sum_{\mathbf{k}^{\prime }\neq \mathbf{k}%
}V_{2D}\left( \mathbf{k}-\mathbf{k}^{\prime }\right) \sin [\theta (\mathbf{k}%
)-\theta (\mathbf{k}^{\prime })]P^{\prime \prime }\left( \mathbf{k}^{\prime
},t\right) .  \label{NPtrF}
\end{equation}%
In Eqs. (\ref{NPRabi}) and (\ref{NPtrF}) $P^{\prime }\left( \mathbf{k}%
,t\right) $ and $P^{\prime \prime }\left( \mathbf{k},t\right) $ are the real
and imaginary parts of $P\left( \mathbf{k},t\right) $, respectively. As is
seen from Eqs. (\ref{ev1})-(\ref{NPtrF}) in the scope of Hartree-Fock
approximation the Coulomb interaction leads to a renormalization of the
light-matter coupling and effective Rabi frequency becomes $\Omega
_{R}\left( \mathbf{k},t\right) +\Omega _{\mathrm{PN}}\left( \mathbf{k}%
,t\right) $. The last term is due to the internal fields and depends on $P$
and $N_{c,v}$. Also, the transition energies become renormalized due to the
Coulomb interaction and we have additional term $\omega _{\mathrm{PN}}\left( 
\mathbf{k},t\right) $. The obtained equations are closed set of nonlinear
integrodifferential equations.

As an initial state we assume undoped TI and for temperature we assume $T<<$ 
$\hbar \omega _{0}$. Hence, for the initial distribution function we take
the limit $T$ $=0$: 
\begin{equation}
N_{v}\left( \mathbf{k},0\right) =1,\ N_{c}\left( \mathbf{k},0\right) =0,\
P\left( \mathbf{k},0\right) =0.  \label{equ}
\end{equation}%
For the initial density matrix (\ref{equ}) (for any isotropic distribution) $%
\Omega _{\mathrm{PN}}\left( \mathbf{k},0\right) =0$ and 
\begin{equation}
\omega _{\mathrm{PN}}\left( \mathbf{k},0\right) =\frac{1}{\hbar \mathcal{A}}%
\sum_{\mathbf{k}^{\prime }\neq \mathbf{k}}V_{2D}\left( \mathbf{k}-\mathbf{k}%
^{\prime }\right) \cos [\theta (\mathbf{k})-\theta (\mathbf{k}^{\prime })].
\label{Self}
\end{equation}%
The latter is the difference of self-energy corrections due to the
electron-electron interactions \cite{Sarma1}, and can be written as 
\begin{equation}
\omega _{\mathrm{PN}}\left( \mathbf{k},0\right) =\frac{e^{2}}{2\pi \hbar
\varepsilon }\int_{0}^{k_{c}}dk^{\prime }\int_{0}^{2\pi }d\theta \frac{\cos
\theta }{\sqrt{k^{2}+k^{\prime 2}-2kk^{\prime }\cos \theta }}.  \label{div}
\end{equation}%
We note that the integral of Eq. (\ref{div}) has an ultraviolet
high-momentum logarithmic divergence, which must be regularized through a
high wave vector cutoff $k_{c}$. As is usual in condensed matter physics,
there is a natural cutoff in the momentum arising from the lattice structure
and, therefore, we have taken $k_{c}=2\pi /a$, where $a=0.41\ \mathrm{nm}$
is the lattice spacing.

Thus the renormalized frequency can be represented as 
\[
\omega _{\mathrm{PN}}\left( \mathbf{k},t\right) =\omega _{\mathrm{PN}}\left(
k,0\right) +\widetilde{\omega }_{\mathrm{PN}}\left( \mathbf{k},t\right) , 
\]%
where $\omega _{\mathrm{PN}}\left( k,0\right) $ is given by the regularized
expression (\ref{div}) and 
\[
\widetilde{\omega }_{\mathrm{PN}}\left( \mathbf{k},t\right) =\frac{2}{\hbar 
\mathcal{A}}\sum_{\mathbf{k}^{\prime }\neq \mathbf{k}}V_{2D}\left( \mathbf{k}%
-\mathbf{k}^{\prime }\right) \sin [\theta (\mathbf{k})-\theta (\mathbf{k}%
^{\prime })]P^{\prime \prime }\left( \mathbf{k}^{\prime },t\right) 
\]%
\begin{equation}
-\frac{2}{\hbar \mathcal{A}}\sum_{\mathbf{k}^{\prime }\neq \mathbf{k}%
}V_{2D}\left( \mathbf{k}-\mathbf{k}^{\prime }\right) \cos [\theta (\mathbf{k}%
)-\theta (\mathbf{k}^{\prime })]N_{c}\left( \mathbf{k}^{\prime },t\right) .
\label{renorm}
\end{equation}%
Because of finite excitation of Brillouin zone around Dirac point now $%
\widetilde{\omega }_{\mathrm{PN}}\left( \mathbf{k},t\right) $ and $\Omega _{%
\mathrm{PN}}\left( \mathbf{k},t\right) $ are convergent. The domain of
integration and the nonlinearity of the light-TI coupling is defined by
dimensionless parameter:%
\begin{equation}
\chi _{0}=\frac{eE_{0}\mathrm{v}_{F}}{\hbar \omega _{0}^{2}},  \label{param}
\end{equation}%
which is the ratio of the amplitude of the momentum given by the wave field
to characteristic excitation momentum $\hbar \omega _{0}/\mathrm{v}_{F}$. In
the limit $\chi _{0}<<1$ the multiphoton effects are suppressed. The
multiphoton effects become essential at $\chi _{0}\sim 1$. To restrict
transitions within the bulk bands one should restrict wave intensity by the
condition%
\begin{equation}
\chi _{0}<<\frac{\mathcal{E}_{g}}{\hbar \omega _{0}}.  \label{tunel}
\end{equation}%
Note that for THz photons the condition (\ref{tunel}) can be fulfilled with
large $\chi _{0}\lesssim 10$.

The terms with partial derivative $\partial /\partial \mathbf{k}$ in the
left-hand side of Eqs. (\ref{ev1})-(\ref{ev3}) describe intraband
transitions. In these equations, we can make a change of variables and
transform the partial differential equation into an ordinary one. The new
variables are $t$ and $\mathbf{k}_{0}=\mathbf{k}-\mathbf{k}_{E}$ $\left(
t\right) $, where 
\[
\hbar \mathbf{k}_{E}\left( t\right) =-e\int_{0}^{t}\mathbf{E}\left(
t^{\prime }\right) dt^{\prime } 
\]%
is the classical momentum given by the wave field.

Equations (\ref{ev1}) and (\ref{ev2}) yield the conservation law for the
particle number:%
\begin{equation}
N_{c}\left( \mathbf{k}_{0},t\right) +N_{v}\left( \mathbf{k}_{0},t\right) =1
\label{cl}
\end{equation}%
With the conservation law (\ref{cl}) one can exclude equation for $%
N_{v}\left( \mathbf{k}_{0},t\right) $.

Note that here we consider a coherent interaction of TI with a pump wave in
the ultrafast excitation regime, which is correct only for the times $t<\tau
_{\min }$, where $\tau _{\min }$ is the minimum of all relaxation times. For
the considered case, at the excitation energies $\mathcal{E}<<\mathcal{E}%
_{g}=0.3\ \mathrm{eV}$, typical for $\mathrm{Bi}_{2}\mathrm{Se}_{3}$, the
dominant mechanism for relaxation will be electron-phonon coupling$\ $via
longitudinal surface phonons \cite{Sarma2,graph}. In the temperature domain $%
2\mathrm{v}_{l}\mathcal{E}/\mathrm{v}_{F}<<T<<\mathcal{E}$, where $\mathrm{v}%
_{l}=2.9\times 10^{5}\ \mathrm{cm/s}$ is the velocity of the longitudinal
acoustic phonon, the relaxation time for the energy level $\mathcal{E}$ can
be estimated as%
\begin{equation}
\tau \left( \mathcal{E}\right) =\left( \frac{D^{2}\mathcal{E}T}{2\rho
_{m}\hbar ^{3}\mathrm{v}_{F}^{2}\mathrm{v}_{l}^{2}}\right) ^{-1}.
\label{rel}
\end{equation}%
Here $D=22\ \mathrm{eV}$ is the deformation potential, and $\rho
_{m}=7.7\times 10^{-7}\ \mathrm{g/cm}^{2}$ is the mass density. For the THz
photon energies $\mathcal{E}=0.004\ \mathrm{eV}$, at temperatures $T=0.1%
\mathcal{E}$ , from Eq. (\ref{rel}) we obtain $\tau \left( \mathcal{E}%
\right) =140\ \mathrm{ps}$. Thus, in this energy range, one can coherently
manipulate with interband multiphoton transitions in TI on time scales $%
t\lesssim 100\ \mathrm{ps}$. For this reason, we consider short pump wave
pulses. The wave amplitude is described by the envelope function $%
E_{0}\left( t\right) =E_{0}f\left( t\right) $:%
\begin{equation}
f\left( t\right) =\left\{ 
\begin{array}{cc}
\sin ^{2}\left( \pi t/\mathcal{T}_{p}\right) , & 0\leq t\leq \mathcal{T}_{p},
\\ 
0, & t<0,t>\mathcal{T}_{p},%
\end{array}%
\right.
\end{equation}%
where $\mathcal{T}_{p}$ characterizes the pulse duration and is chosen to be 
$\mathcal{T}_{p}=32\mathcal{T}$ , where $\mathcal{T}$ $\ $is the wave period.

\section{MULTIPHOTON EXCITATIONS AND GENERATION OF HARMONICS}

The integration of Eqs. (\ref{ev1}), (\ref{ev2}) and (\ref{ev3}) is
performed on a grid of 10000-20000 ($k,\theta $)-points depending on the
intensity of the pump wave. For the integration over polar angle, we use
Gaussian quadrature with $60$ points. For the quantity $k$ we take points
homogeneously distributed between $k=0$ and $k=\alpha \omega _{0}/\mathrm{v}%
_{F}$, where parameter $\alpha $ depends on the intensity of the pump wave.
The time integration is performed with the standard fourth-order Runge-Kutta
algorithm.

The strength of Coulomb interaction is characterized by the dimensionless
parameter $\alpha _{c}$, defined as a ratio of characteristic Coulomb
interaction energy to kinetic energy. For the massless particles, $\alpha
_{c}$ does not depend on the electron density and equals to $\alpha
_{c}=e^{2}/\left( \varepsilon \hbar \mathrm{v}_{F}\right) $. The static
dielectric constant of crystals such as $\mathrm{Bi}_{2}\mathrm{Se}_{3}$ is
estimated to be greater than $50$. We assume that the effective dielectric
constant is the average of that in the TI and in the vacuum, and take a
value of $\varepsilon =20$ \cite{Sarma1}. Thus, for all calculations, we set 
$\alpha _{c}=0.164.$

Photoexcitations of the Fermi-Dirac sea are presented in Figs. 1--3. As a
reference frequency, we have taken $\nu _{0}=\omega _{0}/2\pi =1\ \mathrm{THz%
}$. In Fig. 1 a density plot of the particle distribution function $%
N_{c}\left( \mathbf{k},t_{f}\right) $ after the interaction is shown. The
wave dimensionless amplitude is taken to be $\chi _{0}=0.2$. For this
intensity only one and two-photon transitions take place.

\begin{figure}[tbp]
\includegraphics[width=.5\textwidth]{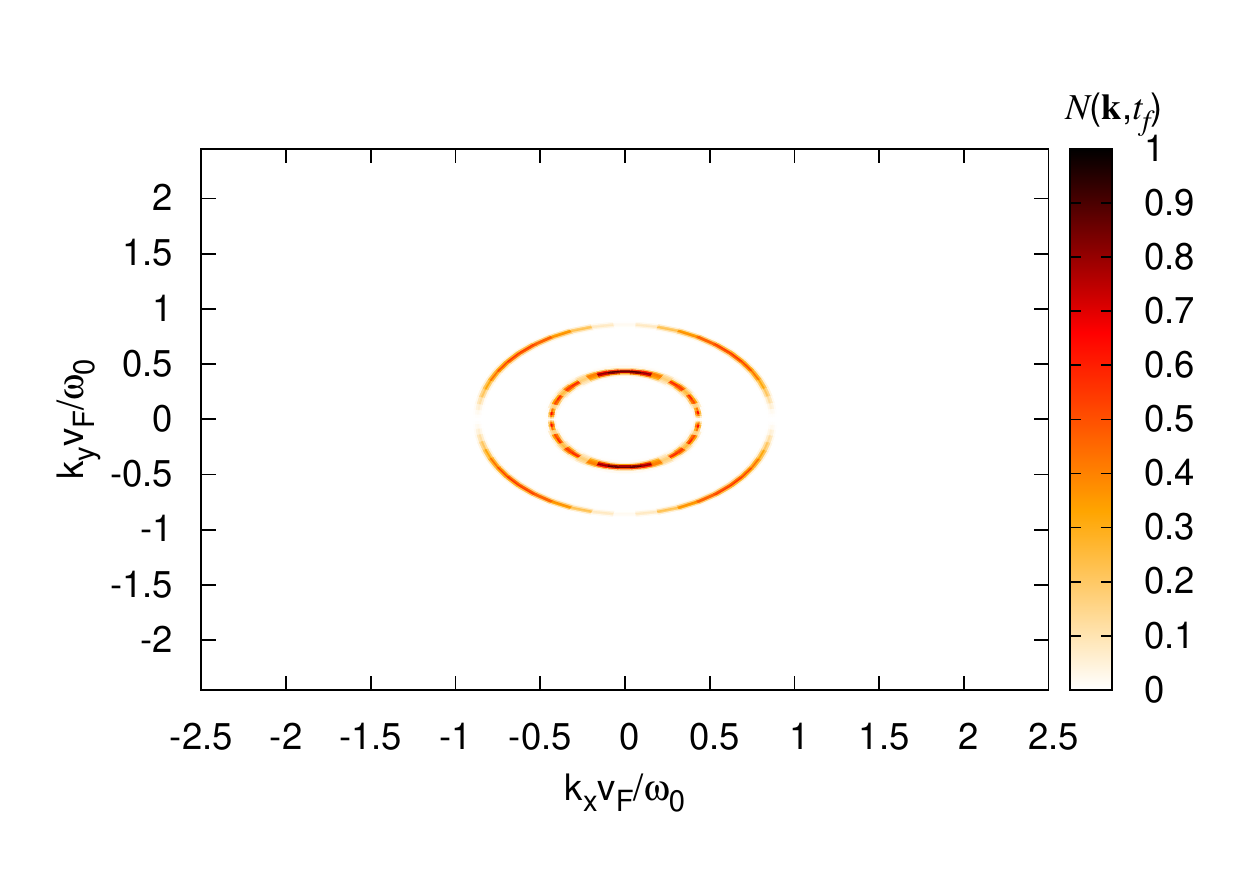}
\caption{(Color online) Particle distribution function $N_{c}\left( \mathbf{k%
},t_{f}\right) $ (in arbitrary units) after the interaction at the instant $%
t_{f}=32\mathcal{T}$, as a function of scaled dimensionless momentum
components. The wave is assumed to be linearly polarized with a
dimensionless parameter $\protect\chi _{0}=0.2$.}
\end{figure}

\begin{figure}[tbp]
\includegraphics[width=.5\textwidth]{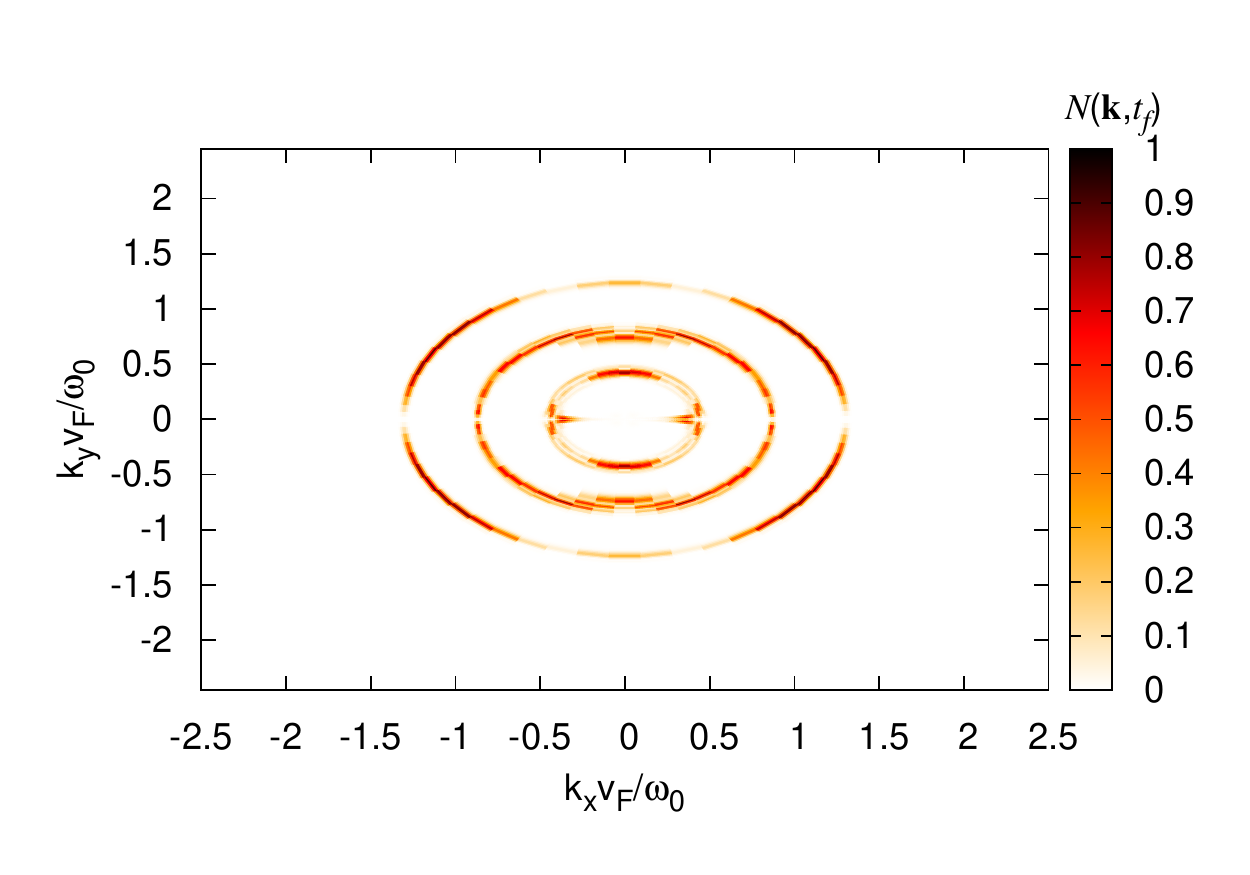}
\caption{(Color online) Creation of a particle-hole pair in TI at
multiphoton excitation. Particle distribution function $N_{c}\left( \mathbf{k%
},t_{f}\right) $ (in arbitrary units) after the interaction is displayed for
a wave intensity $\protect\chi _{0}=0.5$.}
\end{figure}

\begin{figure}[tbp]
\includegraphics[width=.5\textwidth]{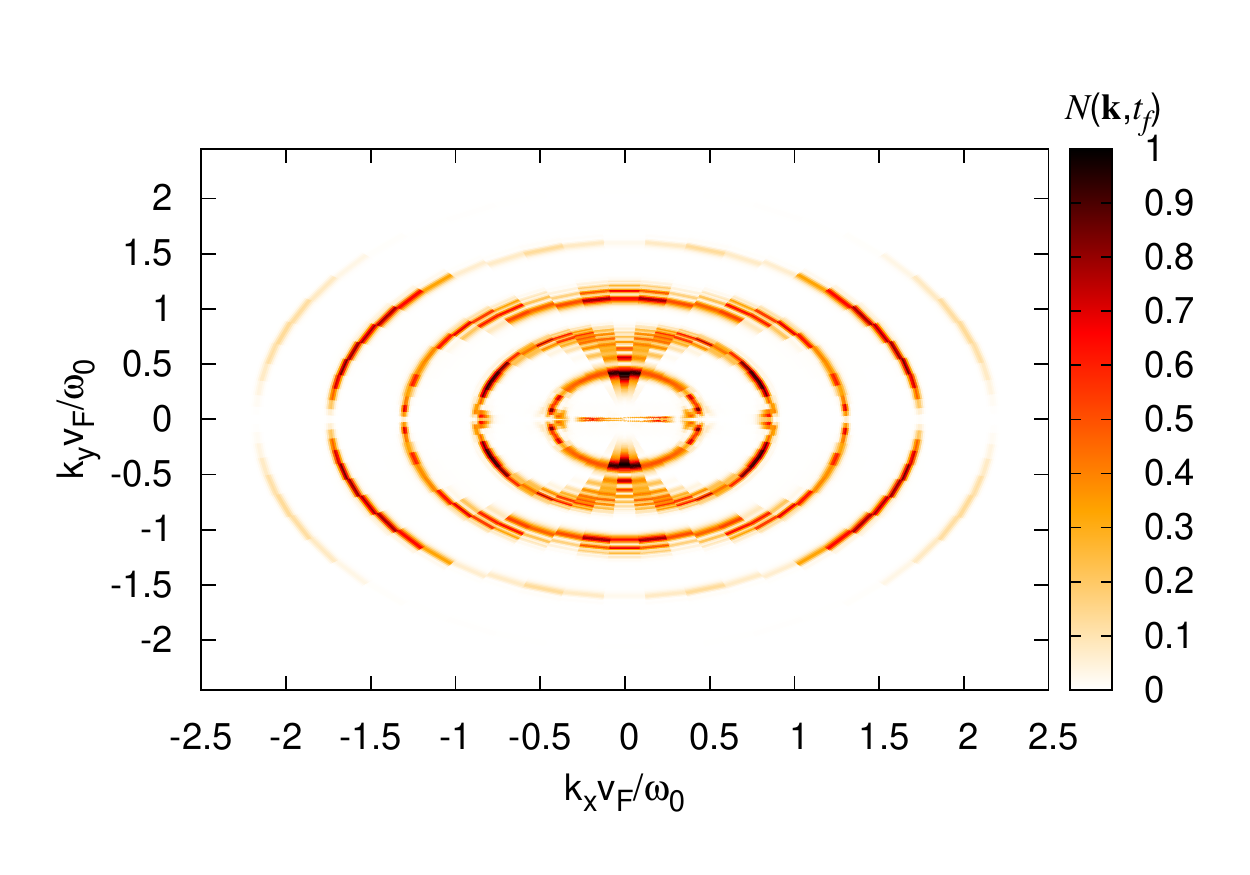}
\caption{(Color online) The same as Fig. 2 but for $\protect\chi _{0}=1$.}
\end{figure}
In Fig. 2 the creation of a particle-hole pair in TI is shown for stronger
wave intensity $\chi _{0}=0.5$. With increasing pump wave intensity and
approaching to the domain $\chi _{0}\sim 1$, the multiphoton excitations
takes place and the Rabi oscillations appear corresponding to multiphoton
transitions. At that, one should take into account the intensity effect of
the pump wave (Stark shift due to free-free intraband transitions) and
Coulomb effect on the quasienergy spectrum. Thus, the multiphoton
probabilities of particle-hole pair production have maximal values for the
resonant transitions%
\begin{equation}
\overline{\omega }\left( \mathbf{k}_{0}\right) =n\hbar \omega _{0},\ \
n=1,2,3...,  \label{rc}
\end{equation}%
where%
\begin{equation}
\overline{\omega }\left( \mathbf{k}_{0}\right) =\frac{1}{\mathcal{T}}%
\int\limits_{0}^{\mathcal{T}}\left( \omega \left( \mathbf{k}_{0}+\mathbf{k}%
_{E}\left( t\right) \right) +\omega _{\mathrm{PN}}\left( \mathbf{k}_{0}+%
\mathbf{k}_{E}\left( t\right) ,t\right) \right) dt  \label{mf}
\end{equation}%
is the mean value of the Coulomb and wave-fields dressed transition
frequency.\ For the effective high-harmonic generation multiphoton
transitions (\ref{rc}) should have reasonable probabilities, that is, the
generalized Rabi frequency and interaction time should be large enough for
full Rabi flopping. As is seen from Fig. 3 at $\chi _{0}=1$, the
probabilities of multiphoton transitions are considerable up to photon
numbers $s=5$. With the multiphoton excitation the total electronic density 
\begin{equation}
n_{c}\left( t\right) =\int N_{c}\left( \mathbf{k},t\right) \frac{d\mathbf{k}%
}{\left( 2\pi \right) ^{2}}  \label{nc}
\end{equation}%
is also varied, approaching to a maximal value, and then falling. The latter
is plotted in Fig. 4. Here $n_{0}=\omega _{0}^{2}/\left( 2\pi \mathrm{v}%
_{F}^{2}\right) $ and for a THz photon $n_{0}=1.43\times 10^{9}\mathrm{cm}%
^{-2}$.

\begin{figure}[tbp]
\includegraphics[width=.48\textwidth]{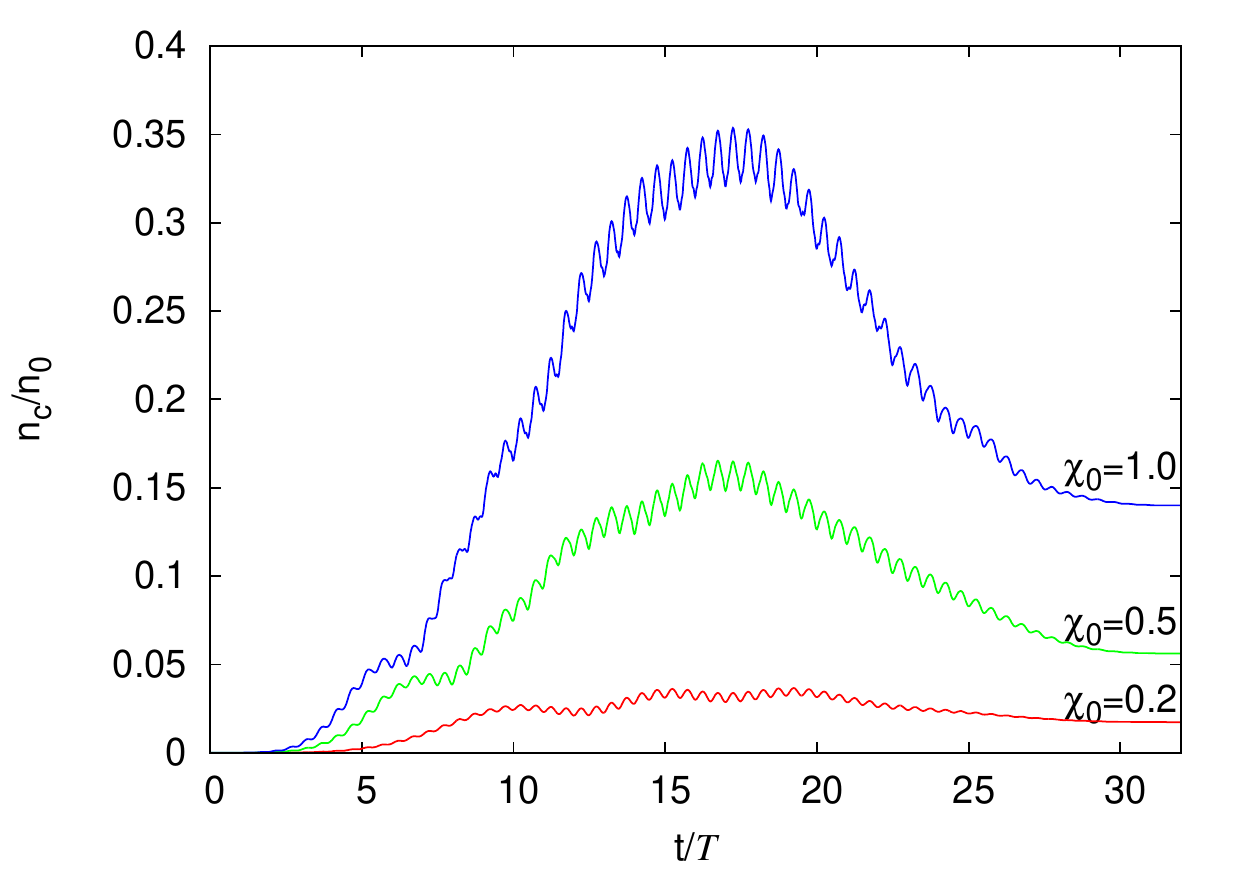}
\caption{(Color online) Time evolution of the total normalized electronic
density (in arbitrary units) at various pump wave intensities.}
\end{figure}
At the multiphoton excitation, the particle-hole annihilation and the
intraband transitions will cause intense coherent radiation of the harmonics
of the applied wave field. Here we consider the possibility of generation of
harmonics at the multiphoton excitation depending on the pump field
intensity and frequency. For the radiation spectrum, one needs the mean
value of the current density operator%
\begin{equation}
\widehat{\mathbf{j}}=-e\langle \ \widehat{\Psi }|\widehat{\mathbf{v}}|\ 
\widehat{\Psi }\rangle ,  \label{jd}
\end{equation}%
where $\widehat{\mathbf{v}}=\hbar ^{-1}\partial H_{0}\left( \mathbf{k}%
\right) /\partial \mathbf{k}$ is the velocity operator. Here we need only
the surface current in the polarization direction of the pump wave: $%
\mathcal{J}_{x}\left( t\right) =\left\langle \widehat{j}_{x}\right\rangle /%
\mathcal{A}$. For the effective Hamiltonian (\ref{DH}) the x-component of
the velocity operator reads%
\begin{equation}
\widehat{\mathrm{v}}_{x}=\mathrm{v}_{F}\left( 
\begin{array}{cc}
0 & -i \\ 
i & 0%
\end{array}%
\right) .  \label{vx}
\end{equation}%
With the help of Eqs. (\ref{exp}), (\ref{def}), (\ref{jd}), and (\ref{vx}),
the surface current can be written as%
\[
\mathcal{J}_{x}\left( t\right) =-\frac{e\mathrm{v}_{F}}{(2\pi )^{2}}\int d%
\mathbf{k}\left[ \cos \theta (\mathbf{k})\left( N_{c}\left( \mathbf{k}%
,t\right) -N_{v}\left( \mathbf{k},t\right) \right) \right. 
\]%
\begin{equation}
\left. +2\sin \theta (\mathbf{k})P^{\prime \prime }\left( \mathbf{k}%
,t\right) \right] .  \label{Jx}
\end{equation}

Thus, having solutions of Eqs. (\ref{ev1}), (\ref{ev2}), and (\ref{ev3}),
then making an integration in Eqs. (\ref{Jx}) one can calculate the harmonic
radiation spectrum with the help of a Fourier transform of the function $%
\mathcal{J}_{x}\left( t\right) $. We assume that the spectrum is measured at
a fixed observation point in the backward propagation direction (and pump
wavelength is much larger than TI film thickness). For the generated field
we have%
\begin{equation}
E_{x}^{(g)}\left( t+z/c\right) =-\frac{4\pi }{c}\mathcal{J}_{x}\left(
t+z/c\right) .  \label{solut}
\end{equation}

The emission strength of the $s$th harmonic will be characterized by the
dimensionless parameter 
\begin{equation}
\chi _{s}=\frac{eE_{x}^{(g)}\left( s\right) \mathrm{v}_{F}}{\hbar \omega
_{0}^{2}}=\chi _{0}\frac{E_{x}^{(g)}\left( s\right) }{E_{0}},  \label{Ks}
\end{equation}%
where%
\begin{equation}
E_{x}^{(g)}\left( s\right) =\frac{\omega _{0}}{2\pi }\int_{0}^{2\pi /\omega
_{0}}E_{x}^{(g)}\left( t\right) e^{is\omega _{0}t}dt  \label{Hs}
\end{equation}%
is the Fourier component of the generated field. In Fig. 5 the density plot
of the radiation spectrum via logarithm of the normalized field strength $%
\chi \left( \omega \right) $ (in arbitrary units) versus the pump wave
intensity is illustrated. Note that with the fast Fourier transform
algorithm instead of discrete functions $\chi _{s}$ we calculate smooth
function$\chi \left( \omega \right) $ and so $\chi _{s}=\chi \left( s\omega
_{0}\right) $. From this figure, we clearly notice maximums at the odd
harmonics and with the increase of the wave intensity the emission strengths
of the high harmonics become feasible and for $\chi _{0}=2$ harmonics up to
21th are sizable. 
\begin{figure}[tbp]
\includegraphics[width=.5\textwidth]{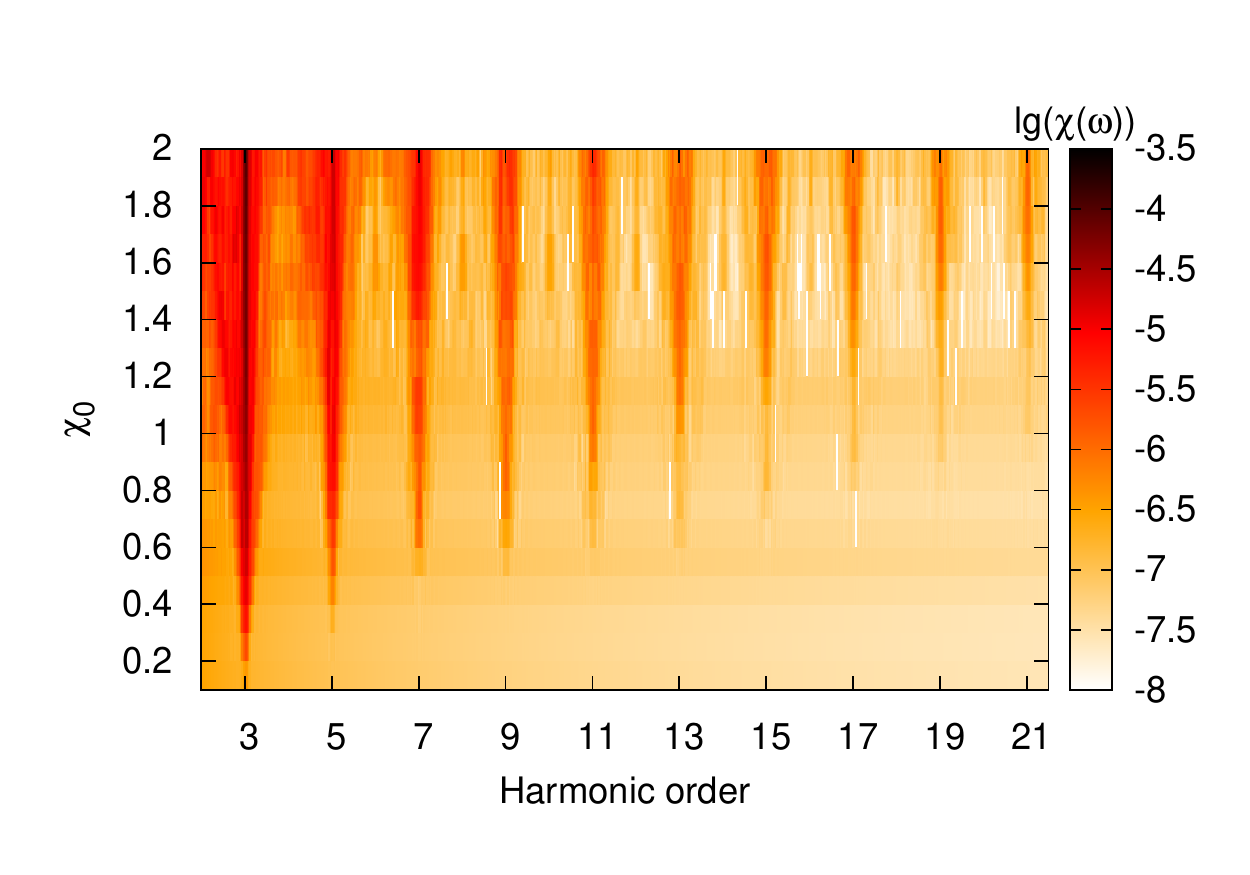}
\caption{(Color online) Density plot of the radiation spectrum via logarithm
of the normalized field strength $\protect\chi \left( \protect\omega \right) 
$ (in arbitrary units) versus the pump wave intensity. The wave frequency is
taken to be $\protect\nu _{0}=1\ \mathrm{THz}$.}
\label{iiii}
\end{figure}

We further examine emission rates of the 3rd and 5th harmonics for various
pump wave frequencies versus intensity, which are shown in Figs. 6 and 7.
For the considered intensities the perturbation theory is not applicable and
in Figs. 6 and 7 we have a strong deviation from power law for the emission
rates of harmonics. In particular, the rate of the 3rd harmonic scales is
almost linearly on the pump wave strength $\chi _{3}\sim \chi _{0}$. Whereas
it should show the $\chi _{0}^{3}$ dependence in the perturbative limit.
Besides, these figures show that the emission rates almost independent of
the pump wave frequency. 
\begin{figure}[tbp]
\includegraphics[width=.46\textwidth]{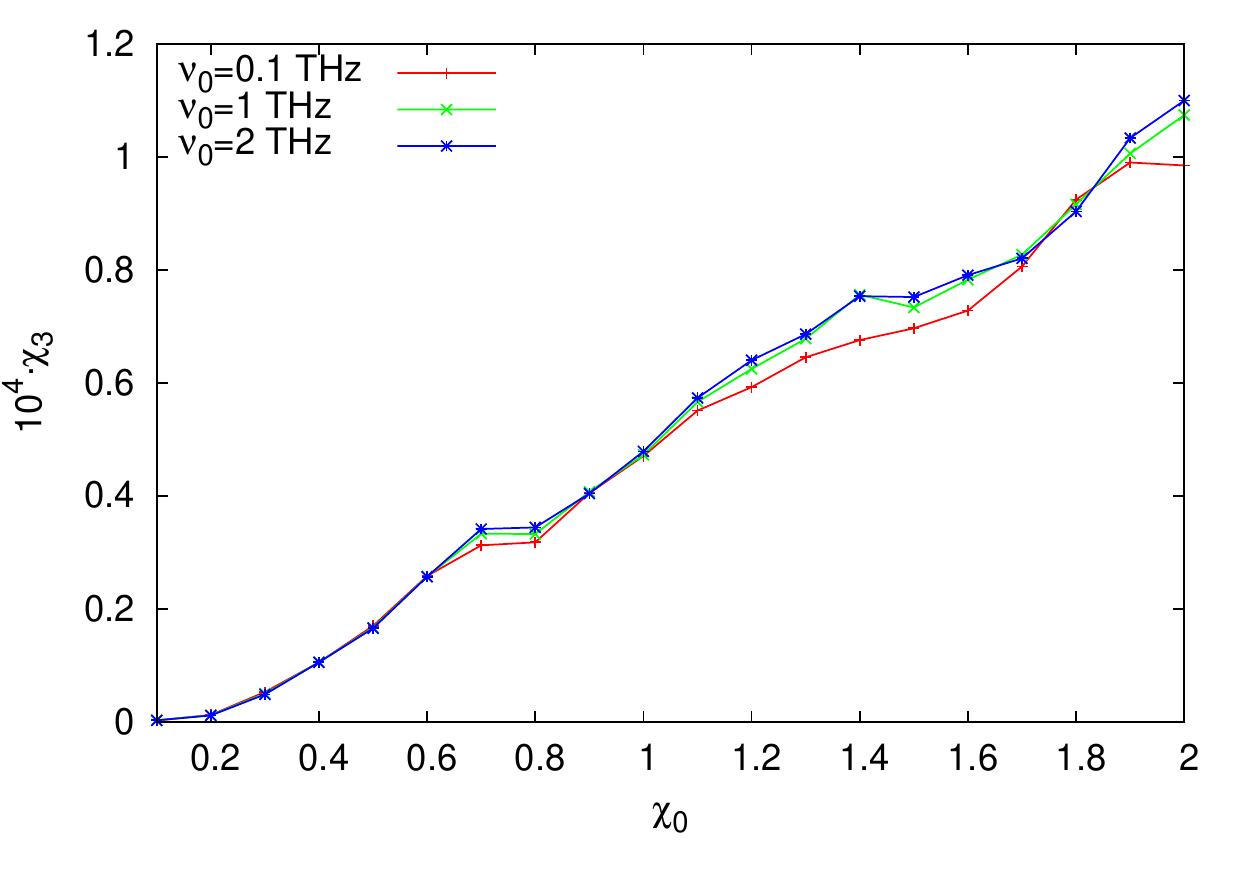}
\caption{(Color online) Third harmonic emission rate in TI versus pump wave
intensity for various wave frequencies.}
\end{figure}
\begin{figure}[tbp]
\includegraphics[width=.46\textwidth]{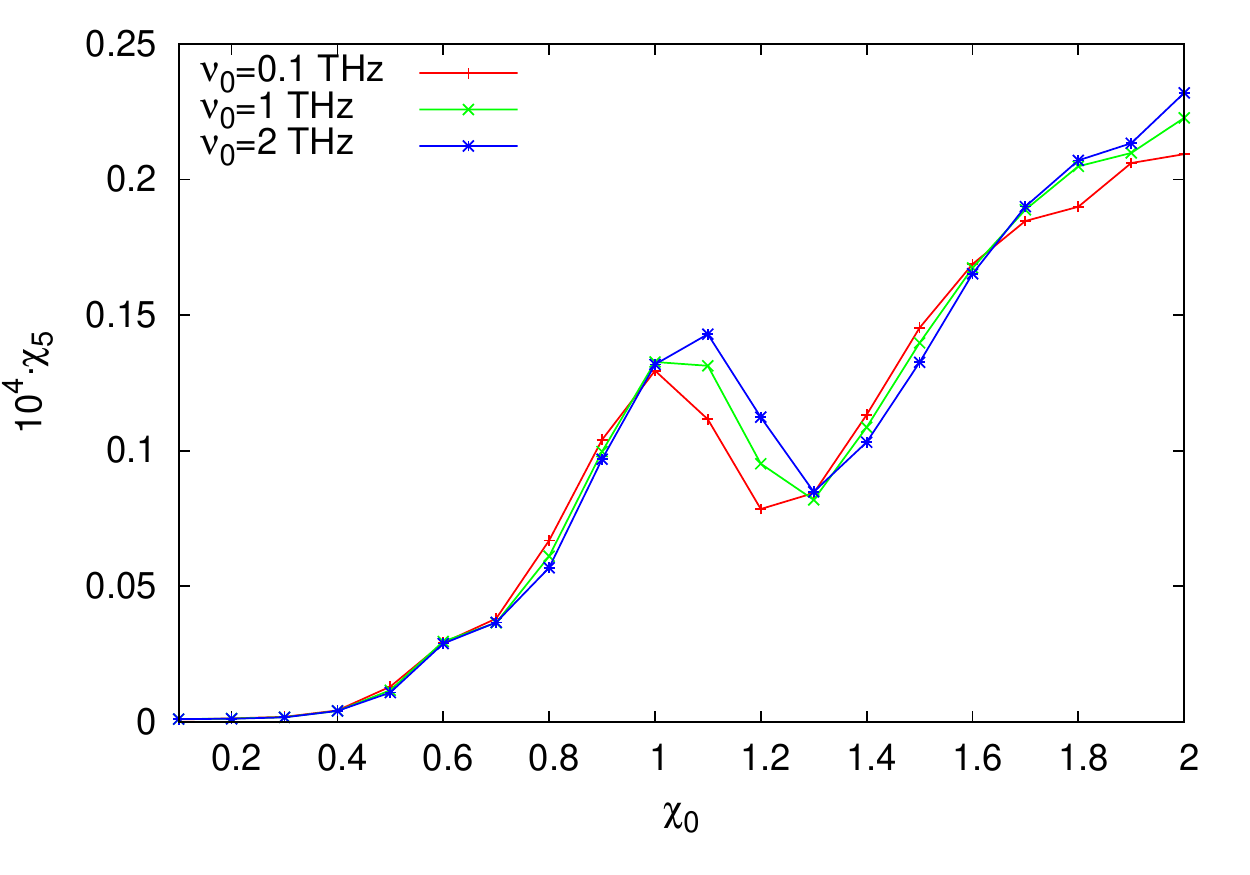}
\caption{(Color online) Fifth harmonic emission rate in TI versus pump wave
intensity for various wave frequencies.}
\end{figure}
Thus, calculations show that at the multiphoton excitation of 2D metallic
surface states of TI the generation of high harmonics is possible which
takes place for the wide range of pump wave frequencies. The average
intensity of the wave expressed by $\chi _{0}$, can be estimated as 
\begin{equation}
I_{\chi _{0}}=\chi _{0}^{2}\times 2\times 10^{2}\ \mathrm{W\ cm}^{-2}\left( 
\frac{\nu _{0}}{\mathrm{THz}}\right) ^{4}.  \label{int}
\end{equation}

The intensity $I_{\chi _{0}}$ strongly depends on the pump wave frequency.
In particular for THz pump waves, high-harmonics can be generated at the
intensities $I_{\chi _{0}\approx 1}\approx 2\times 10^{2}\ \mathrm{W\ cm}%
^{-2}$.

\section{Conclusion}

We have presented a nonlinear microscopic theory of the TI interaction with
coherent electromagnetic radiation in the ultrafast excitation regime.
Electron-electron Coulomb interaction has been taken into account with the
self-consistent Hartree-Fock approximation that leads to a closed set of
integrodifferential equations for the interband polarization and carrier
occupation distribution. The dynamics of the multiphoton excitation of 2D
metallic surface states of TI depending on the wave intensity has been
considered and analyzed on the basis of numerical simulations. It has been
shown that by THz radiation of moderate intensities, one can control
interband multiphoton transitions in 2D metallic surface states of TI on
time scales $t\lesssim 100\ \mathrm{ps}$. Furthermore, we have shown that
along with multiphoton transitions there is an intense radiation of high
harmonics at the interband (particle-hole annihilation) and intraband
transitions induced by a pump wave. The obtained results certify that the
process of high-harmonic generation for THz photons can be already observed
for intensities $\sim 0.2\ \mathrm{kW\ cm}^{-2}$ and temperatures $T<<$ $%
\hbar \omega _{0}$.

This work was supported by the RA MES State Committee of Science and
Belarusian Republican Foundation for Fundamental Research (RB) in the frames
of the joint research projects SCS AB16-19 and BRFFR F17ARM-25, accordingly.


\section*{References}

\begin{thebibliography}{99}
\bibitem{TIR1} Hasan M Z, Kane C L 2010 \textit{Rev. Mod. Phys.} \textbf{82}
3045

\bibitem{TIR2} Qi X L, Zhang S C 2011 \textit{Rev. Mod. Phys.} \textbf{83}
1057

\bibitem{exp1} Hsieh D, Xia Y, Qian D, Wray L, Meier F, Dil J H, Osterwalder
J, Patthey L, Fedorov AV, Lin H, Bansil A \textit{Phys. Rev. Lett. }2009 
\textbf{103} 146401

\bibitem{exp2} Zhang T, Cheng P, Chen X, Jia J F, Ma X, He K, Wang L, Zhang
H, Dai X, Fang Z, Xie X 2009 \textit{Phys. Rev. Lett.} \textbf{103} 266803

\bibitem{TIL2} Tse W K, Macdonald A H 2010 \textit{Phys. Rev. Lett.} \textbf{%
105} 057401

\bibitem{TIL3} Tse W K, Macdonald A H 2010 \textit{Phys. Rev. B} \textbf{82}
161104

\bibitem{TIL4} Hosur P 2011 \textit{Phys. Rev. B} \textbf{83} 035309

\bibitem{PhotD} Iurov A, Gumbs G, Roslyak O, Huang D 2013 \textit{J. Phys.:
Condens. Matter} \textbf{25} 135502

\bibitem{Magnetao} Rahim K, Ullah A, Tahir M, Sabeeh K 2017 \textit{J.
Phys.: Condens. Matter} \textbf{29} 425304

\bibitem{Kerr} Jenkins G S, Sushkov A B, Schmadel D C, Butch N P, Syers P,
Paglione J, Drew H D 2010 \textit{Phys. Rev. B} \textbf{82} 125120

\bibitem{Farad} Sushkov A B, Jenkins G S, Schmadel D C, Butch N P, Paglione
J, Drew H D 2010 \textit{Phys. Rev. B} \textbf{82} 125110

\bibitem{2nd} Hsieh D, McIver J W, Torchinsky D H, Gardner D R, Lee Y S,
Gedik N 2011 \textit{Phys. Rev. Lett.} \textbf{106} 057401

\bibitem{DiracM} Peres N M R, Santos J E 2013 \textit{J. Phys.: Condens.
Matter} \textbf{25} 305801

\bibitem{THzpl} Autore M, Di Pietro P, Di Gaspare A, D'Apuzzo F, Giorgianni
F, Brahlek M, Koirala N, Oh S, Lupi S 2017 \textit{J. Phys.: Condens. Matter}
\textbf{29} 183002

\bibitem{THzStrong} Giorgianni F, Chiadroni E, Rovere A, Cestelli-Guidi M,
Perucchi A, Bellaveglia M, Castellano M, Di Giovenale D, Di Pirro G,
Ferrario M, Pompili R 2016 \textit{Nat. Commun.} \textbf{7} 11421

\bibitem{HH2} Mikhailov S A, Ziegler K 2008 \textit{J. Phys.: Condens. Matter%
} \textbf{20} 384204

\bibitem{Mer1} Avetissian H K, Avetissian A K, Mkrtchian G F, Sedrakian Kh V
2012 \textit{Phys. Rev. B} \textbf{85} 115443

\bibitem{Mer2} Avetissian H K, Avetissian A K, Mkrtchian G F, Sedrakian Kh V
2012 \textit{J. Nanophoton.} \textbf{6}, 061702

\bibitem{Mer3} Avetissian H K, Ghazaryan A G, Mkrtchian G F, Sedrakian Kh V
2017 \textit{J. Nanophoton.} \textbf{11} 016004

\bibitem{HH3-exp} Yoshikawa N, Tamaya T, Tanaka K 2017 \textit{Science} 
\textbf{356} 736

\bibitem{Sarma1} Abergel D S L, Das Sarma S 2013 \textit{Phys. Rev. B} 
\textbf{87} 041407

\bibitem{Sarma2} Das Sarma S, Li Q 2013 \textit{Phys. Rev. B} \textbf{88}
081404

\bibitem{graph} Viljas J K, Heikkil\"{a} T T 2010 \textit{Phys. Rev. B} 
\textbf{81}, 245404
\end{thebibliography}
\end{document}